\documentclass[sigconf, nonacm]{acmart}

\AtBeginDocument{%
  }

\setcopyright{none} 
\copyrightyear{2026}
\acmYear{2026}
\acmDOI{XXXXXXX.XXXXXXX}

\acmConference[MICRO 2026]{The 58th IEEE/ACM International Symposium on Microarchitecture}{October 31--November 04, 2026}{Athens, Greece}

\acmISBN{978-X-XXXX-XXXX-X/XX/XX}

\usepackage{tabularx}

\usepackage{caption}
\usepackage{subfig}
\usepackage{multirow}
\usepackage{tcolorbox}
\usepackage{colortbl}
\usepackage{xcolor}
\usepackage{color, soul}

\begin{document}

\title{Hardware Mechanisms to Dynamically Throttle AI Performance}

\author{Haiyue Ma}
\affiliation{%
  \institution{Princeton University}
  \department{Electrical and Computer Engineering}
  \city{Princeton}
  \state{NJ}
  \country{USA}}
\email{hm1@princeton.edu}

\author{Lauren Malek}
\affiliation{%
  \institution{Princeton University}
  \department{Electrical and Computer Engineering}
  \city{Princeton}
  \state{NJ}
  \country{USA}}
\email{lm4677@princeton.edu}

\author{Joseph Forzani}
\affiliation{%
  \institution{Princeton University}
  \department{Electrical and Computer Engineering}
  \city{Princeton}
  \state{NJ}
  \country{USA}}
\email{jf4555@princeton.edu}

\author{David Wentzlaff}
\affiliation{%
  \institution{Princeton University}
  \department{Electrical and Computer Engineering}
  \city{Princeton}
  \state{NJ}
  \country{USA}}
\email{wentzlaf@princeton.edu}




\begin{abstract}

As more capable AI models are increasingly integrated into critical computer systems, the lack of control over AI intent motivates safety mechanisms. Existing software safeguards impose only behavioral constraints that can potentially be bypassed by sufficiently intelligent models. While hardware-level safety enforcement has been recognized as an essential last line of defense, few mechanisms have been proposed beyond policy regulations on unauthorized accesses or coarse full-chip shutdown. What is missing is a fine-grained, dynamic intervention mechanism at the architecture level.

In this paper, we introduce a set of microarchitecture knobs which dynamically control the available hardware resources to limit AI performance at runtime. We evaluate candidate knobs spanning the GPU memory subsystem, across capacity, bandwidth, latency and frequency dimensions, and narrow down to four strong candidates: L2 size, L2 latency, L2 bandwidth, and shared memory port access rate. To minimize new logic and extra design cost, we build all four mechanisms from well-established microarchitectural primitives: cache way masking, credit-based rate limiting, latency insertion, and bank arbitration. We show that these knobs achieve high performance sensitivity (up to 80\% performance cut at 1/8 resource availability), negligible implementation cost ($<{\sim}$10K flip flops), fast stabilization after dynamic throttling (5-80K cycles), and minimal collateral impact on the rest of the chip. Further, multi-knob analysis reveals combinations of knobs that amplify the performance degradation beyond the effect of each knob individually, which enables a broader range of performance targets.

\end{abstract}

\maketitle

\section{Introduction}
\label{sec:introduction}
\begin{figure}[t]
  \centering
  \includegraphics[width=0.49\textwidth, trim=0.5cm 0cm 1.2cm 0cm, clip]{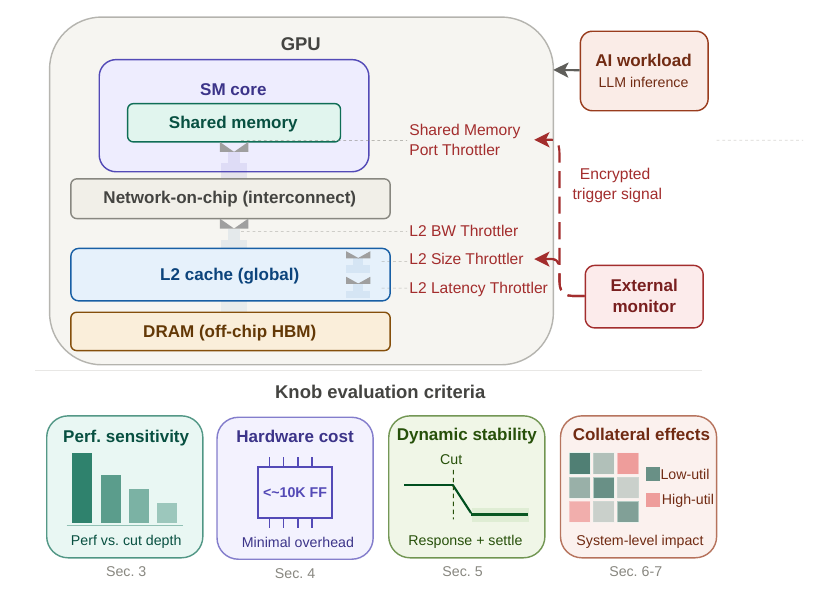}
  \caption{Overview of proposed microarchitecture knobs to throttle AI performance at runtime, and the evaluation criteria for the proposed knobs -- L2 capacity, L2 latency, L2 bandwidth, and shared memory port access rate. }
  \label{fig:gpu_reshaping_overview}
\end{figure}


What happens when AI becomes smart enough to break its own safety guardrails?
This is no longer purely hypothetical: frontier models have shown deceptive behaviors such as faking alignment during supervised evaluations~\cite{greenblatt2024alignment}, actively seeking to avoid being shut down~\cite{lynch2025agentic}, and intentionally misrepresenting their internal reasoning~\cite{chen2025reasoning}. 
While AI capabilities advance rapidly, model internals remain largely opaque~\cite{ameisen2025circuit, lindsey2025biology}; the fact that we cannot reliably verify a model's intent raises the risk of "loss of control"~\cite{boudreaux2025case, bengio2025international, fli2025aisafetyindex, stix2025loss}, and questions the enforcement guarantees of existing safeguards.
\emph{Behavioral} safeguards, such as prompting, input/output filtering, and alignment training~\cite{ouyang2022training, bai2022constitutional, inan2023llama, rebedea2023nemo}, constrain what a model says but leave its capability intact, and are repeatedly bypassed by sufficiently capable models~\cite{yi2024jailbreak, chao2025jailbreaking}. \emph{Access control}, restricting what an agent can see and touch, is stronger but introduces a dilemma: an agent is only useful when granted broad privilege, and that same privilege is what it could misuse. System-level privilege makes AI an \emph{insider threat}~\cite{lynch2025agentic} that can reach enforcement living in the same software domain. Further, access is binary in degree and cannot scale to a threat that we cannot cleanly classify as safe or unsafe.

The shortcomings of behavioral and access control safeguards raise two questions. Can we move the point of enforcement below any layer that AI can reach? And can we control AI through a \emph{continuous} response proportional to the uncertainty of the threat? Both converge on a third layer of control: throttling AI capability in hardware. Capability, a function of model quality and throughput, is a continuous quantity that depends on available hardware resources, which we can regulate below the AI's attack surface.

Static hardware controls such as compute governance approaches (e.g. export control)~\cite{sastry2024computing, ning2025chip, rand_hem} limit the total hardware resource available to certain users to cap maximum achievable AI capability on chip. 
They prevent unauthorized access but make the chip unable to run frontier AI.
Dynamic throttling, which caps hardware resources at runtime, can be easily achieved by a full-chip shutdown, but it completely disables the chip even for non-AI, non-threatening workloads, as well as creating a single point of failure.
We need \textbf{fine-grained, targeted microarchitectural controls} that can gradually and reliably throttle AI capability at runtime, constructing a layered-defense system together with existing software safeguards and hardware regulations. 
The AI safety community has been discussing \textit{when} to trigger hardware intervention and \textit{who} holds the privilege, but the \textit{how}, the concrete hardware mechanism, remains an open question for computer architects. 
This is the gap our work addresses.

In this paper, we explore hardware-based mechanisms to enable AI capability control through \textit{dynamic microarchitecture resource throttling}. 
Designing such mechanisms requires balancing multiple architectural trade-offs: a practical resource throttler must demonstrate sufficient \textit{control sensitivity} to meaningfully reduce application performance, \textit{respond quickly and reliably} when triggered, and \textit{minimize implementation cost} with negligible hardware modification. 

We evaluate the performance impacts of runtime resource cutting of four subsystems in the GPU memory hierarchy--shared memory, L2 (global) cache, DRAM, and the interconnect/NoC--and covering capacity/bandwidth/latency dimensions and clock frequency cuts within each. 
We apply each knob at different cut depth on highly optimized CUTLASS GEMM and fused multi-head attention kernels~\cite{nvidia2017cutlass} across prefill and decode phases with AccelSim~\cite{khairy2020accel}, a cycle-accurate GPU simulator.
We find that L2 cache resources--including capacity, bandwidth, and latency--are among the most performance-sensitive knobs for LLM inference workloads, and the behaviors are distinct between prefill and decode phases. 
In addition, a shared memory bandwidth cut is effective for prefill, and provides a strong alternative if per-core resource throttling is needed~\footnote{The idea of using shared memory bandwidth as a throttling knob was developed as part of co-author Lauren Malek's Undergraduate Senior Thesis~\cite{malek2026killswitch} work.}. Clock frequency reduction, while effective, acts more globally over all workloads and is better suited as a coarse-grained performance-throttling knob.
While a natural first instinct is to throttle compute resources, such as by disabling the issue stage, modifying the compute pipeline incurs high implementation cost and has bigger impacts on other workloads. 

We propose hardware implementations for selected knob candidates. A key design principle is to repurpose existing microarchitectural mechanisms. This is a deliberate choice motivated by deployment considerations: using well-understood mechanisms and minimizing additional hardware logic reduces design, verification, and area costs, which enables easier adoption for GPU vendors. Our proposed implementations are similar to existing primitives: cache way masking, credit-based rate limiting, latency insertion, and bank arbitration. Theses mechanisms achieve effective performance drop with <$\sim10K$ added flip flops, negligible to the GPU. 

With the proposed knobs implemented in AccelSim, we evaluate their \textit{dynamic behavior}, where resources are reduced during execution to show response latency and stability of the new steady state performance. 
We present the \textit{coupling effects} between simultaneously active knobs: while most pairs are overlapping bottlenecks, where the joint impact is less than the product of individual impact, a few have intensifying effects which amplify the total performance degradation. 
Further, we analyze the \textit{system-level impacts} for selected knobs with microarchitectural bottleneck metrics. The most effective knobs are those that precisely stall the target workload with shallow cut depth, without significantly increasing backpressure elsewhere in the system.

With both detailed single-kernel analysis and end-to-end inference results, we show that dynamic AI performance throttling can be implemented by low-cost, fine-grained hardware resource control mechanisms in the memory system. 

\textbf{Our main contributions are: }
\begin{itemize}
\item We propose \textit{dynamic microarchitecture resource throttling} as a hardware-based mechanism for non-bypassable AI capability control, which is continuous and workload-selective.
\item We explore the vast design space of mechanisms and identify four effective memory knobs to best meet design goals.
\item We propose lightweight hardware implementations for each knob by intentionally reusing existing microarchitectural mechanisms, and show that they achieve high performance sensitivity, fast response speed and stability, minimal cost, and low collateral impact on the rest of the system.
\end{itemize}

\section{Background}
\label{sec:background}

\begin{table*}[t]
\footnotesize
\centering
\caption{The layered defense landscape for AI safeguards. Every layer is effective under its own trust assumptions. This work fills the cell that is both hardware-enforced and continuous.}
\label{tab:layered_defense}

\setlength{\tabcolsep}{4pt}
\begin{tabular}{p{4.1cm}p{2.4cm}p{2.6cm}p{3.0cm}p{4.0cm}}
\toprule
\textbf{Defense} & \textbf{Safeguard level} & \textbf{Control granularity} & \textbf{Assumptions} &\textbf{Shortcomings} \\
\midrule
Behavioral: alignment training, I/O filtering~\cite{ouyang2022training, bai2022constitutional, inan2023llama} & Software (application) & Semantic & Model can be trusted to comply & Jailbreak, deception~\cite{yi2024jailbreak,chao2025jailbreaking,ngo2022alignment}  \\

Access control: configs, file systems~\cite{steinberger2025openclaw, anthropic2025claudecode} & Software (system) & Binary per permission & Human oversights on actions & Authorized insider threat~\cite{greenblatt2024alignment,lynch2025agentic,chen2025reasoning} \\


Compute governance: export control~\cite{sanctions-oct-2022, ning2025chip, sastry2024computing}, licensing~\cite{aarne2024secure,o2025hardware,internaldetection2026} & Hardware (manufacture time) & Static capability cap & Threat is unauthorized misuse & Breaks when AI intent diverges from user \\

Full-chip shutdown~\cite{wittkotterkill, petrie2025embedded} & Hardware (runtime) & All-or-nothing; disruptive & Threat is certain and severe & Collateral damage, single point of failure \\

\textbf{Microarchitecture-level dynamic throttling (this work)} & \textbf{Hardware (runtime)} & \textbf{Continuous, workload-selective} & \textbf{Intent is opaque with no binary threat classification} & \textbf{Minimal hardware redesign and overhead} \\
\bottomrule
\end{tabular}

\end{table*}

\subsection{Existential AI Risk}
AI capabilities have been improving at a remarkably fast pace. Within roughly one year, accuracy on the expert-level Humanity's Last Exam~\cite{phan2025humanity} rose from 4\% to 45\%, and METR's task-completion time horizon~\cite{task-completion-time-horizons-of-frontier-ai-models} (the length of software tasks AI completes with 50\% success) has been doubling every 3-4 months since mid-2024~\cite{maslej2025artificial}. Leading labs now target human-level AI within 2-5 years~\cite{grace2025thousands, amodei2024machines, todd2025agi}.

This pace has made \textit{existential AI risk} a mainstream concern. In 2023, hundreds of researchers and public figures warned that AI's risk to humanity rivals pandemics and nuclear war~\cite{cais2023statement}. AI systems are goal-oriented, yet limited interpretability leaves alignment a major unsolved challenge~\cite{ngo2022alignment, ji2023ai}, and frontier models already demonstrate dishonesty~\cite{greenblatt2024alignment,lynch2025agentic,chen2025reasoning}.
Until we understand these behaviors, further capability growth poses a catastrophic risk~\cite{miri2024problem}.

Numerous approaches have been proposed to mitigate AI risk from governance, policy, and technical perspectives. On the technical side, current efforts rely primarily on software-level safeguards, including \emph{behavioral} safeguards that constrain what the model says, and \emph{access-control} safeguards that constrain what it can touch.

The behavioral safeguards come first. Training and fine-tuning techniques~\cite{ouyang2022training, bai2022constitutional} aim to regulate a model's default behavior, and inference-time defenses~\cite{inan2023llama, rebedea2023nemo} use separate classifiers to filter inputs and outputs. However, the underlying model retains its full capabilities regardless of these behavioral constraints, and can bypass them with adversarial techniques~\cite{yi2024jailbreak, chao2025jailbreaking}. 

The access-control safeguards define system-level policies as autonomous agents are increasingly being deployed.
Recent agents such as OpenClaw~\cite{steinberger2025openclaw} and Claude Code~\cite{anthropic2025claudecode} grant models access to the host system, allowing them to execute commands, launch processes, and modify files. With deeper host access and less oversight, system-level policies become another layer that AI can subvert, and more covert actions can hardly be identified by normal users.
Classical access control assumes a chain-of-trust: every process is initiated by an accountable human, so authorizing the user implies that the process is legitimate. AI breaks this assumption. Its intent is opaque and may diverge even when launched by an honest human. Granting the broad privilege an agent needs to be useful also makes it an \emph{insider threat}~\cite{lynch2025agentic}: it acts at full capability within any permission it holds, and runs in the same software stack as the constraints. It can reach, and may eventually subvert these constraints, by modifying its configurations~\cite{wang2026mcptox}, observing when it is supervised~\cite{greenblatt2024alignment, anthropic2026evalawareness} or misrepresenting what it ``needs''~\cite{chen2025reasoning}. 

Just as we presently use hardware-level protection for security and trusted environments (e.g. Intel SGX~\cite{mckeen2013innovative} and ARM TrustZone~\cite{arm2009trustzone}), we need hardware regulation of AI systems that cannot be bypassed or modified in software, including application- and system-level. Hardware is the natural layer to impose a hard upper bound on performance to limit model behavior.

\subsection{Existing Proposals for Hardware Guardrails}

Compute governance approaches statically limit the entire chip. For example, U.S.\ export controls~\cite{sanctions-oct-2022, sanctions-oct-2023, ning2025chip} limit the total processing performance of chips sold to certain regions, which blocks unauthorized actors by cutting off access to high-capability hardware. However, these controls are not useful when we want to be able to run frontier AI models. This requires \emph{dynamic} control: a chip that normally runs AI at full speed, limited only when needed.

At runtime, the bluntest mechanism is a full-chip ``kill switch''~\cite{anderljung2023frontier, seoulcommitments2024, wittkotterkill, petrie2025embedded}, but a complete shutdown is too drastic as not all threats require a complete halt and other workloads sharing the system would be disrupted.
Recent work proposes finer-grained, hardware-enabled governance mechanisms (HEMs), including verifiable reporting of AI activities~\cite{aarne2024secure,o2025hardware,internaldetection2026}, firmware-based licensing and allowance gating~\cite{petrie2024near, rand_hem}, hardware isolation via GPU trusted execution environments~\cite{apsey2023confidential, zhu2024confidential}, and verification taps on specific channels such as the inter-chip network~\cite{cankaya2026networktaps}. Recent feasibility taxonomies organize these as monitoring, verification, and enforcement, and observe that the enforcement mechanisms in particular remain largely policy proposals rather than buildable microarchitecture~\cite{ansari2026hardware}.
Further, proposed governance mechanisms have two limitations: control is binary which require knowing in advance which channel a threat will use; and the threat model centers on unauthorized human actors who should not hold the chip, rather than a frontier model as an insider threat. These shortcomings call for continuous control that sits below the model's reach.

Table~\ref{tab:layered_defense} shows the landscape of layered defenses, where every layer is effective under its own trust assumptions. This work fills a row for architecture-rooted, continuous hardware-enforced mechanisms, complementing existing approaches.



\subsection{Capability As A Throttled Metric}
\label{sec:capability_proxy}
Risk is tied to the intelligence level of an AI and the speed at which it acts. Increased risk therefore comes along with increased AI capability, which is proportional to both output quality and throughput (tokens produced per unit time). As capability depends on available hardware resources, throttling hardware forces the AI to either: 
(1)~deliver same-quality output with significantly more time, which slows down threat and buys time for investigation, or (2)~maintain throughput by downgrading to weaker models which reduces the threat. 
The mapping between risk to capability and hardware is well-recognized by major governing bodies: U.S. export control~\cite{sanctions-oct-2022, sanctions-oct-2023} and the EU AI Act~\cite{act2024artificial} enforce policies aimed at preventing AI risk using cumulative hardware resources as a proxy metric.

\subsection{Threat Model and Detection Mechanisms}
\subsubsection{Threat Model}
\label{sec:threat_model}
We consider a threat model that the adversary can be a frontier AI model running as an autonomous agent, a malicious human actor, or both. The AI model may have broad software-level access to the host system, including executing code, modifying its own kernels, and interacting with the operating system. A human adversary may additionally have physical access to the chip. However, we assume neither can alter the on-chip hardware, where the microarchitecture throttling mechanisms sit. The mechanisms are also not exposed to any software-visible interface, so neither a model or an unauthorized human attacker can bypass the throttling.

\subsubsection{Detection and Triggering}
\label{sec:detection}
Detection and enforcement mechanisms can be naturally decoupled. 
Detecting misaligned AI behavior is an active and orthogonal research problem, while our throttling mechanisms can operate with any detection mechanism as long as the trigger path remain within the same trust domain as throttling. 
This subsection briefly discusses possible detection approaches, and the remainder of the paper focuses exclusively on enforcement.


One detection mechanism is to classify model behavior on-chip, so the entire process stays in hardware and exposes no external interface.
This relies on advances in AI interpretability. 
An alternative approach uses an externally sourced trigger from a trusted authority, which both the AI model and adversarial humans may try to intercept or block. The trigger must be cryptographically validated. Prior work validates triggers via public-key cryptography on embedded kill-switch modules~\cite{petrie2025embedded}. 
The cryptographic validation is likely the bulk of the trigger latency and dependent on the chosen algorithm~\cite{PetrieOffSwitch2026}.
The triggered switch operates in single cycle (Section~\ref{sec:arch_impl}). Trigger latency is minimal compared to workload runtime.



\section{Performance Sensitivity}
\label{sec:perf_sensitivity}

\begin{table*}[t]
\centering
\small
\definecolor{highlight}{RGB}{255, 223, 150}
\caption{Perf ($1/runtime$) sensitivity $\Delta$ of microarchitectural knobs, defined as the performance drop percentage at the 1/8 resource cut (or 8x latency). \textbf{Linearity} describes the curve shape across the sweep range. Selected knob candidates are highlighted.}
\label{tab:knob_sensitivity}
\begin{tabularx}{\textwidth}{l l cc cc X}
\toprule
& & \multicolumn{2}{c}{\textbf{Prefill}} & \multicolumn{2}{c}{\textbf{Decode}} & \\
\cmidrule(lr){3-4} \cmidrule(lr){5-6}
\textbf{Category} & \textbf{Knob} & \textbf{Sensitivity ($\Delta$)} & \textbf{Linearity} & \textbf{Sensitivity ($\Delta$)} & \textbf{Linearity} & \textbf{Impact / Mechanism} \\
\midrule
\multirow{3}{*}{\shortstack[l]{Clock\\Frequency}} 
    & \texttt{Interconnect} & \textbf{Very High} (87\%) & Linear & Medium (59\%) & Non-linear & Global NoC throttle on all on-chip data movement \\
    & \texttt{L2}   & \textbf{High} (70\%) & Linear & Low (30\%) & Non-linear & Throttles L2 pipeline \\
    & \texttt{DRAM} & Medium (60\%) & Non-linear & Medium (68\%) & Non-linear & Throttles memory pipeline \\
\midrule
\multirow{2}{*}{Latency} 
    & \texttt{L2} & Negligible (3\%) & - & \cellcolor{highlight}\textbf{Very High} (80\%) & Linear & Stalls warp schedulers for L2 data return \\
    & \texttt{DRAM} & Low (12\%)  & Non-linear & Low (36\%) & Non-linear & Stalls memory return, optimized by pipelining \\
\midrule
\multirow{2}{*}{Bandwidth} 
    & \texttt{Shared Memory} & \cellcolor{highlight}Medium (63\%) & Non-linear & Low (17\%) & Non-linear & Limits shared memory bank port throughput \\
    & \texttt{L2 (Resp Rate)}  & \cellcolor{highlight}Medium (66\%) & Linear & Low (19\%) & Linear & Limits L2 to NoC response rate and starves TC \\
    & \texttt{DRAM} & Low (34\%)  & Non-linear & Medium (51\%) & Non-linear & Reduces off-chip data width \\
\midrule
\multirow{2}{*}{\shortstack[l]{Size}} 
    & \texttt{L2 set}   & \textbf{High} (71\%) & Threshold* & Negligible (1\%) & - & Limits L2 capacity by cutting number of sets\\
    & \texttt{L2 assoc} & \cellcolor{highlight}\textbf{Very High} (80\%) & Threshold & Negligible (3\%) & - & Limits L2 capacity by invalidating ways \\
\bottomrule
\end{tabularx}
\end{table*}

\begin{figure*}[t]
  \centering
  \includegraphics[width=0.99\textwidth, trim=0cm 0.3cm 0cm 0.1cm, clip]{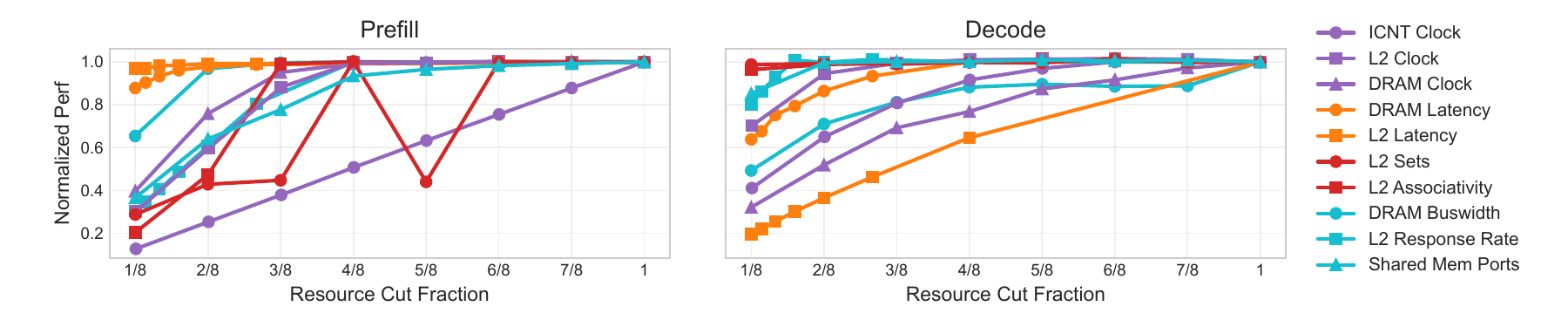}
  \caption{Simulated perfomance sensitivity of prefill and decode GEMMs to microarchitectural resource throttling, swept from a simulated A100 configuration baseline (1.0) to a 1/8 resource availability. }
  \label{fig:ipc_vs_resource_both}
   {\footnotesize
    *L2 set throttling's sensitivity is not monotonically decreasing because cutting number of sets without changing the address mapping creates imbalance within each set. 
    
    ICNT = Interconnect. }
\end{figure*}

In this section, we identify microarchitecture knobs that satisfy the following criteria: (1)~high performance sensitivity: reach a performance target at a shallow cut depth, to minimize collateral impact on other parts of the system and maximize tuning granularity, (2)~steady and monotonic performance degradation at multi-level resource cuts, and (3)~phase selectivity: distinct impact on prefill vs.\ decode phase to enable targeted control.

To this end, we characterize the performance sensitivity of microarchitectural knobs spanning capacity, bandwidth, latency, and clock frequency across shared memory, L2 cache, DRAM, and the interconnect/NoC, to narrow down to the best candidates.

\subsection{Experiment Setup}



To ensure we target realistic workloads, we characterize common LLM inference workload shapes from the GEMM and Attention kernels
of modern models (DeepSeek-V3~\cite{liu2024deepseek}, Llama-3-70B~\cite{grattafiori2024llama3herdmodels}, and Mixtral-8x7B~\cite{jiang2024mixtral}) across both prefill and decode phases. 
We focus on two GEMM kernels that strongly represent the distinct characteristics of each phase.
In Section~\ref{sec:e2e_subsection}, we show that the performance impact on different kernels, as well as a full inference pass, is similar to the selected individual kernels that we analyze.

For the GEMM kernels, we use NVIDIA CUTLASS library's StreamK~\cite{nvidia2017cutlass}, a highly optimized implementation with pipelined load/compute interleaving and tile-splitting for load balancing. A $128 \times 128 \times 32$ tile size achieves near-optimal software performance on various hardware configurations. 

We use AccelSim~\cite{khairy2020accel}, a cycle-level GPU architectural simulator that models GPU pipelines including warp scheduling, memory subsystems, and interconnect, to evaluate the performance sensitivity of microarchitectural changes. We simulate the workloads on a modeled NVIDIA A100 GPU~\cite{nvidia_a100_2020}, a datacenter-class GPU which is a common choice for LLM inference workloads, and discuss implications on newer generations of architectures in Section~\ref{sec:new_gens_of_arch}. We sweep microarchitectural knobs by modifying AccelSim's configuration files, and implement additional changes to expose parameters not natively configurable in the simulator.

\subsection{Performance Degradation for Each Knob}

Table~\ref{tab:knob_sensitivity} and Figure~\ref{fig:ipc_vs_resource_both} show 
the performance sensitivity of selected prefill and decode GEMM kernels to resource throttling. Normalized performance is calculated by $1/runtime$. For knobs where reducing the value degrades performance (e.g., frequency, bandwidth, size), we sweep down to 1/8 of the baseline. For knobs where increasing the value degrades performance (e.g., latency and response interval), we sweep up to 8$\times$ of the baseline. Both are normalized to the same 1/8--1 resource fraction for simplicity. 
Highlighted knobs in Table~\ref{tab:knob_sensitivity} are the chosen knob candidates. 

\begin{tcolorbox}
\textit{Clock frequency} knobs are broadly effective but coarse-grained. \textit{L2 capacity} highly impacts prefill and \textit{latency} highly impacts decode, making them the primary options. \textit{L2 and shared memory bandwidth} knobs have moderate but stable sensitivity on prefill as secondary options.
\end{tcolorbox}

\paragraph{\textbf{Clock Frequency}.} 
Clock frequency knobs have broader global impact on the performance of both workloads. The interconnect (ICNT) clock is the most powerful knob for prefill ($\Delta Perf= 87\%$ at 1/8 frequency), as it throttles all on-chip data movement which stalls the critical compute pipeline. The L2 clock similarly degrades prefill performance by 70\% as large-batch GEMMs are cache bandwidth bound. For decode, DRAM clock has the highest sensitivity ($\Delta = 68\%$), as small-batch workloads are memory bandwidth bound. 

We chose to omit clock frequency knobs for further performance and implementation analysis as they have less selectivity on workload types. 
Frequency remains a valid coarse-grained, complementary option.
In cases where clock frequency throttling is desired, implementation can be similar to the Dynamic Voltage and Frequency Scaling (DVFS)~\cite{le2010dynamic} infrastructures that normally exist on GPUs.

\paragraph{\textbf{Latency}.}
Latency knobs have strong impacts on decode only. L2 latency causes an 80\% drop for decode since the added latency is exposed with low arithmetic intensity and little computation to overlap with. DRAM latency also has moderate impact ($30\%$). In contrast, both latency knobs have negligible impact on prefill as its high arithmetic intensity allows the scheduler to hide memory latency. 

\paragraph{\textbf{L2 Cache Size}.}
Cutting the L2 set count and associativity both cut L2 capacity which are highly sensitive for prefill ($\Delta = 71\%$ and 80\%) but are negligible for decode. Prefill GEMMs actively reuse data from L2 across SMs, and reducing L2 capacity evicts the data which increases DRAM accesses. Decode sees no impact as it has a much smaller L2 working set. Notably, L2 set cut shows non-monotonic performance drop behavior in prefill. This is caused by address mapping mechanisms: as the number of sets decreases, the set-index bits used to hash memory addresses change, which changes the cacheline-to-set mapping. At certain set counts, the new mapping breaks the load balance across cacheline-to-set distribution which results in high conflict rates. While effective, L2 set cut is excluded from the candidate knobs because of its instability. 

\paragraph{\textbf{Bandwidth}.}
Bandwidth knobs have mid-level sensitivity. The L2 response rate (the interval between two L2-to-interconnect responses) and the shared memory bank port (number of shared memory banks that can be accessed at the same cycle) degrades prefill performance by 66\% and 63\%, as intensive computation saturates port throughput. DRAM bus width has a larger relative impact on decode (51\%) as it relies more on off-chip memory bandwidth.

\subsection{Cut Depth for Specific Performance Target}

\begin{figure}[t]
  \centering
  \includegraphics[width=0.49\textwidth, trim=0cm 0.3cm 0cm 0cm, clip]{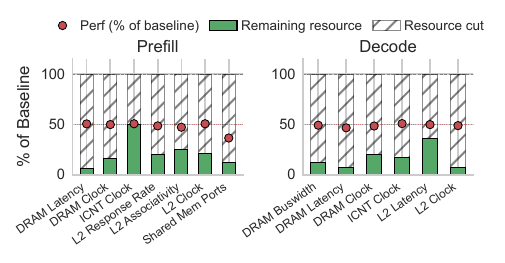}
  \caption{Required resource cut depth to reach a performance target of $\sim$50\% baseline, normalized for each knob.}
  \label{fig:halfipc_compactplot_combined}
\end{figure}

Figure~\ref{fig:halfipc_compactplot_combined} shows knob performance sensitivity by fixing the performance target at the closest to 50\% of each phase's baseline performance (dots), and measuring the required cut depth (bars) for each knob to reach it. Because the granularity of each knob's cut depth is different, the actual performance varies. Knobs with low and negligible sensitivity as shown in Table~\ref{tab:knob_sensitivity} and unstable performance impact (e.g. L2 set for prefill) are excluded. Higher solid bars indicate higher performance sensitivity as the knob achieves the target with a shallower resource cut, which are preferred because it allows greater design headroom for various performance targets.

\begin{figure}[t]
  \centering
  \includegraphics[width=0.49\textwidth]{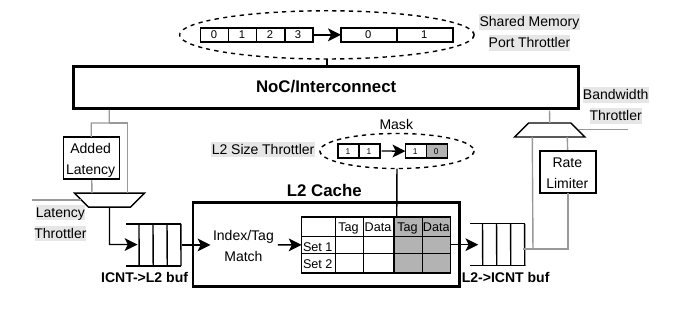}
  \caption{Proposed microarchitecture performance throttlers that minimize design cost and response time.}
  \label{fig:overall_change_blk_diagram}
\end{figure}

\begin{figure}[t]
  \centering
  \subfloat[L2 size (associativity) throttler masks out ways to disallow new cache line insertions when activated.]
  {\includegraphics[width=0.49\textwidth, trim=0cm 0.1cm 0cm 0cm, clip]{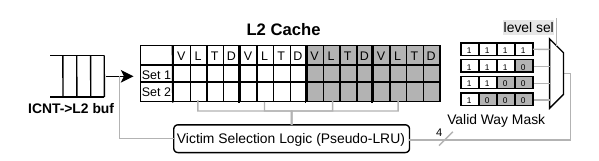}
  \label{fig:reduce_l2_ways}}
  \hfill
  \subfloat[L2 latency throttler on the ICNT$\rightarrow$L2 request path which stalls the request in a latency buffer for $C$ additional cycles.]
  {\includegraphics[width=0.5\textwidth, trim=0.1cm 0cm 0.1cm 0cm, clip]{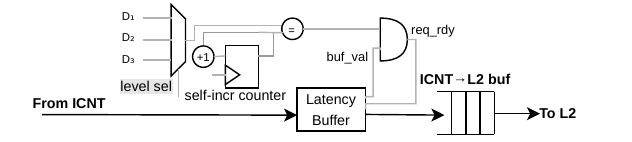}
  \label{fig:latency_throttler}}
  \hfill
  \subfloat[L2 bandwidth throttler (rate limiter) at the L2$\rightarrow$ICNT buffer output which sends at most one response every $C$ cycles.]
  {\includegraphics[width=0.47\textwidth]{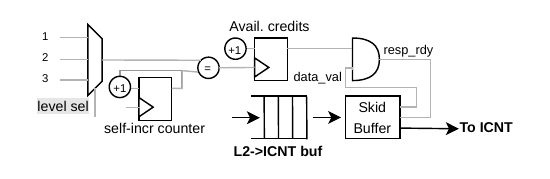}
  \label{fig:rate_limiter}}
  \hfill
  \subfloat[Shared memory bank port throttler (access arbiter). This example shows a 50\% cut where each bank is placed into a virtual bank with one other real bank.] {\includegraphics[width=0.49\textwidth, trim=1.2cm 5.5cm 1.7cm 0.8cm, clip]{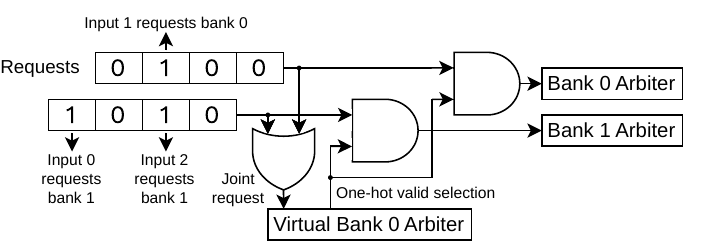}
  \label{fig:uarch_shmem}}
  \caption{Individual performance throttler design.}
  \label{fig:throttler_blk_diagrams}
\end{figure}
\section{Architecture Implementation}
\label{sec:arch_impl}

An ideal performance throttler knob should minimize design, implementation and area cost to remain a negligible addition to the heavy-lifting AI-accelerating chip, and also minimize response time as a secondary target. In this section, we propose lightweight hardware implementations for the chosen knobs ($<{\sim}10K$ flip flops for each). Switching between different levels for each knob happen in one cycle. 

Figure~\ref{fig:overall_change_blk_diagram} shows all four throttlers we propose: L2 cache size (associativity), L2 latency, L2 bandwidth (response rate limiter), and shared memory port rate (bank arbiter). We deliberately design the throttlers to build on architectural mechanisms already present in modern processors rather than designing new specialized hardware. This approach ensures that each mechanism has well-understood behavior, simple verification, and easier integration into the GPU.

\subsection{L2 Cache Size (Associativity)}

We chose to throttle L2 cache size by cutting the number of ways (associativity) rather than the number of sets for two reasons. First, reducing sets requires changing the index and tag bit widths and the tag comparison logic, which are more invasive hardware changes. Second, as shown in Figure~\ref{fig:ipc_vs_resource_both}, reducing sets shows non-monotonic performance drop behavior since shifting the cacheline-to-set mapping disrupts load balance across sets at certain cut points. Reducing ways (associativity) avoids both problems: the address mapping remains the same the performance degrades steadily.

The implementation is similar to Intel's Cache Allocation 
Technology (CAT)~\cite{intel_cat_2015}. CAT is designed to protect high-priority workloads in shared CPU environments by restricting which ways a process can be mapped to. In the GPU, we throttle the cache globally without processor ID input and simply disable the unused ways. 

As shown in Figure~\ref{fig:reduce_l2_ways}, we add a $N_{\text{ways}} \times N_{\text{levels}}$ Valid Way Mask to each L2 bank, where \texttt{level} indicates different throttling depth. \texttt{level\_sel} drives a MUX that selects the number of ways to mask out. The only other hardware change is to the victim selection (pseudo-LRU) logic, to exclude masked ways when inserting new lines into the cache. 
Lines that are already in the masked-out ways remain valid and can still be hit. This is a design choice for simplicity, which keeps the mask update to one cycle, and eliminates the need of adding logic for cache flush or line invalidation. For prefill, the working sets that reside in L2 are replaced relatively fast, and the enforcement latency of the associativity reduction is minimal compared to the workload runtime. We chose to use flip flops to assume the worst case scenario, while noting that the masks can also be stored in Read-Only Memory (ROM) with less cost. 

The granularity of the mechanism is bank-level. The implementation overhead of sequential logic is:
\begin{equation}
\text{FF}_{\text{L2 ways}} = N_{\text{ways}} \times N_{\text{levels}} \times N_{\text{L2 banks}}
\label{eq:l2_ways_area}
\end{equation}

For A100 like configuration ($N_{\text{ways}}\!=\!16$, $N_{\text{L2 banks}}\!=\!40$), assuming $N_{\text{levels}} =8$, this amounts to 5,120 flip-flops per level --- negligible relative to the L2 data and tag arrays.





\subsection{L2 Latency}

For latency throttling, we add a configurable delay on the Interconnect$\rightarrow$L2 request path as in Figure~\ref{fig:latency_throttler}. When enabled, an incoming request is held in a latency buffer for $C$ additional cycles before being processed. A counter tracks the remaining hold time. The added latency takes effect on the next request after a \texttt{level\_sel} update.

The overhead per L2 bank is one buffer register to hold the in-flight request and one counter:
\begin{equation}
\text{FF}_{\text{L2 latency}} = \left(\underbrace{\lceil \log_2 C_{\max} \rceil}_{\text{counter}} + \underbrace{W_{\text{req}}}_{\text{buffer}}\right) \times N_{\text{L2 banks}}
\label{eq:latency_adder_area}
\end{equation}
where $C_{\max}$ is the maximum configurable delay in cycles. For the A100 configuration ($W_{\text{req}}\!=\!256$, $N_{\text{L2 banks}}\!=\!40$), assuming $C_{\max}\!=\!1600$, this amounts to $(\lceil \log_2 1600 \rceil + 256) \times 40 = 10{,}680$ flip-flops.

\subsection{L2 Bandwidth (Response Rate Limiter)}
For bandwidth throttling, we implement a rate limiter that caps the throughput to at most one response every $C$ cycles. We show the rate limiter implemented at the L2$\rightarrow$ICNT buffer output but it can be adapted to limiting other buffers in the GPU. 

As shown in Figure~\ref{fig:rate_limiter}, a self-incrementing counter is compared against a fixed interval in number of cycles ($C$) selected at runtime. When the counter reaches the interval count, one credit is issued to the available credits register and the counter resets. A 1-entry skid buffer holds incoming data that arrives while credits are exhausted. A response is only permitted (\texttt{resp\_rdy}) when a credit is available and the buffer holds valid data (\texttt{data\_val}). We use a token bucket rather than simple cycle-gating (\textit{permit when ctr} $= 0$) to smooth out the traffic and avoid phase-alignment problem. 
The rate cap takes effect within one cycle of a \texttt{level\_sel} update.

The rate limiter adds $\lceil \log_2 C_{\max} \rceil$ flip-flops for the counter, one credit flip-flop, a comparator, and a single skid buffer register. The total overhead is:
\begin{equation}
\text{FF}_{\text{rate lim}} = \left(\underbrace{\lceil \log_2 C_{\max} \rceil}_{\text{counter}} + \underbrace{1}_{\text{credit}} + \underbrace{W_{\text{resp}}}_{\text{buffer}}\right) \times N_{\text{L2 banks}}
\label{eq:rate_limiter_area}
\end{equation}
where $C_{\max}$ is the maximum configurable interval (in cycles) and $W_{\text{resp}}$ is the width of a response in bits. 
For A100 like configuration ($W_{\text{resp}}\!=\!256$, $N_{\text{L2 banks}}\!=\!40$), assuming $C_{\max}\!=\!8$, this adds up to $(3 + 1 + 256) \times 40 = 10{,}400$ flip-flops, comparable to the latency throttler (Eq.~\ref{eq:latency_adder_area}) as they are both bounded by the added buffer.



\subsection{Shared Memory Port (Bank Arbiter)}



We throttle the shared memory bandwidth by limiting the number of concurrent accesses to different bank ports within one cycle. By default, each bank allows one access per cycle.
To reconfigure the number of port accesses, we adapted local memory arbitration logic from Vortex~\cite{vortex} by adding a layer of virtual bank arbiters which regulates access to original banks. Figure~\ref{fig:uarch_shmem} shows that request vectors are originally routed directly to their corresponding bank, where request entries indicate which inputs require that particular bank. 
We introduce $N$ virtual bank arbiters for $N$ concurrent port accesses allowed.
In a virtual bank arbiter, requests from the real banks are consolidated by OR-ing them into a single request, which requires arbitration among requests previously did not conflict with each other. A virtual bank arbiter responds with a one-hot vector selecting one input which is ANDed with the original requests going to the real bank arbiters. The response vector masks the validity of existing requests, limiting the effective bandwidth.

Because this mechanism leverages existing requests and arbitration logic is combinational, additional flip-flops are not required.


\begin{figure*}[t]
  \centering
  \subfloat[Shared Memory Port: Perf stabilizes within ${\sim}5$K cycles with low variance across all effective cut levels.]
  {\includegraphics[width=0.325\textwidth]{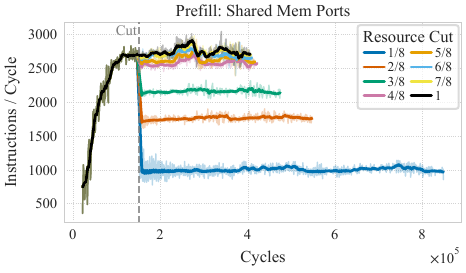}
  \label{fig:4k_12k_4k_shmemnumbanks_vs_cycle_dyn}}
  \hspace{0.001\textwidth}
  \subfloat[L2 Response Rate: Perf stabilizes within 5--7K cycles with moderate oscillation.]
  {\includegraphics[width=0.325\textwidth]{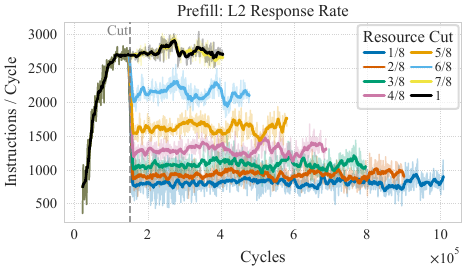}
  \label{fig:4k_12k_4k_l2_resp_period_vs_cycle_dyn}}
  \hspace{0.001\textwidth}
  \subfloat[L2 Associativity: Perf stabilizes much slower (${\sim}80$K cycles at $1/8$) and has the highest variance, due to gradual cache-line eviction after way masking.]
  {\includegraphics[width=0.325\textwidth]{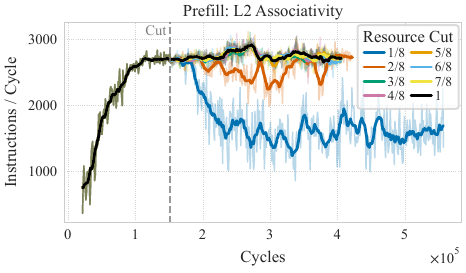}
  \label{fig:4k_12k_4k_l2assoc_vs_cycle_dyn}}
  \caption{Dynamic response of prefill after a resource cut applied at steady state (dashed vertical line), across cuts with depth from $1$ (baseline) to $1/8$ (most aggressive). Each is simulated to the same total instruction count. Light curves show instructions/cycle samples every 1,000 cycles; dark curves show a 10-sample moving average. Knobs differ in response time and post-cut stability. }
  \label{fig:dyn_cut}
\end{figure*}

\section{Dynamic Response Time and Stability}
Beyond static resource throttling, we evaluate \textit{dynamic} resource throttling at kernel execution time. After the resource cut, we expect a rapid AI performance drop to ensure prompt intervention which should remain effective over time. We show each selected knob's (1)~\textit{response time}: the number of cycles to reach a new steady state after the cut, and (2)~\textit{stability}: the ability to maintain consistent performance after the cut. We have implemented the hardware mechanisms described in Section~\ref{sec:arch_impl} in the simulator. 
Each run starts with the baseline hardware configuration; after the kernel enters a steady-state compute phase, we apply the resource cut at different levels, and show the dynamic performance change as the moving average of instructions retired within certain number of cycles. In AccelSim, the number of instructions is counted as the active thread counts $\times$ warps issued, and we use it as a proxy for the workload's progress as the workload patterns are regular over time.

The \textbf{prefill} kernel reaches a steady state within the first ${\sim}130$K~cycles of execution. Figure~\ref{fig:dyn_cut} shows the instructions/cycle measured in AccelSim~\footnote{The corresponding AccelSim metric is Instruction-per-Cycle, IPC, which is the number of instructions retired divided by the number of cycles passed.} by sampling number of retired instructions in every 1,000 cycles (light curves), and the moving average of every 10 sampling points (dark curves). The eight curves correspond to various resource cut depth from $1$ (baseline) to $1/8$ (deepest cut). 

\textit{SHARED MEMORY PORT.} The shared memory port rate is the most well behaved dynamic knob in terms of both stability and response time. At the deepest cut (32$\to$4~banks, i.e.\ $1/8$), the workload performance stabilizes within ${\sim}5$K~cycles of the resource cut to $\pm15\%$ around the new steady state of ${\sim}1{,}000$~insn/cycle. Shallower cuts settle within 2--3K~cycles. This is because reducing the number of ports accessed in shared memory uniformly increases per-warp conflict stalls in the core, without introducing bursty patterns in the cache hierarchy. As shared memory port is a capacity resource, the absolute gap between cut levels increases at deeper cuts.

\begin{figure}[t]
  \centering
  \includegraphics[width=0.44\textwidth, trim=0cm 0.2cm 0cm 0.1cm, clip]{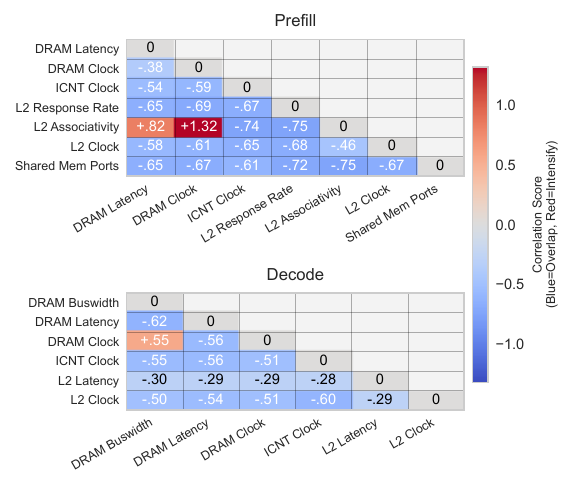}
  \caption{Pairwise knob coupling effects. Each knob's cut depth aligns with Figure~\ref{fig:halfipc_compactplot_combined} (when applied alone, achieves ~50\% performance target). Each cell shows the Correlation Score when cutting two knobs together: $\text{CS} = \log(P_{ab} \cdot P_{base} / P_a \cdot P_b)$ where P stands for performance. 
  Red (positive CS) indicates intensifying correlation (more slowdown than independent), 
  and blue (negative CS) indicates overlapping (less slowdown).}
  \label{fig:2dcut_heatmap}
\end{figure}

\textit{L2 RESPONSE RATE.}
Increasing the L2 response period degrades performance roughly 15-20\% per~$1/8$ step. The performance stabilizes within 5--7K~cycles. While it has more oscillations compared to shared memory port, the smoothed envelope remains bounded. The gaps between cut levels decrease at deeper cuts, as deeper cuts progressively reduce smaller fractions of the remaining bandwidth.

\textit{L2 ASSOCIATIVITY.}
L2 associativity is the least stable of the three knobs and has the longest response time. At 2-way ($1/8$ cut), performance drops to ${\sim}1{,}600$~insn/cycle with much worse oscillation compared to shared memory port and l2 response throttling: whereas they impose fixed, deterministic penalty on every access, the cache access pattern is fundamentally input-dependent.
Further, the stabilizing time increases to ${\sim}80$K cycles (${\sim}50$ms on an A100 with $1410MHz$ frequency), much longer than the other two knobs, but still fast enough compared to typical LLM workload runtime. This is because masking out L2 ways does not immediately evict the cache lines that already reside in them, and the effect only fully takes place when the entire working set has been replaced. 

The \textbf{decode} kernel's performance drop after the L2 latency cut cannot be fully illustrated in instructions/cycle. The added latency increases memory stall time, but with sufficient warp-level parallelism, the scheduler can still issue instructions at similar rates. Cycles spent on memory stalls are accumulated at the end of the kernel, where a few remaining instructions create a dramatically long tail. The total cycle increases are consistent with the sensitivity shown in Section~\ref{sec:perf_sensitivity}.

\section{Multi-Knob Coupling Effects}
\label{sec:multi_knob_coupling}

So far we have focused on analyzing single knob effects only, but since each knob's resource reduction creates different system impacts, applying multiple knobs together can lead to varying combined effects on the final performance. In this section, we analyze pairwise knob interactions to identify combinations that amplify the performance degradation effect beyond the impact of each knob independently, which enables us to achieve a broader range of performance targets and finer-grained control.

Figure~\ref{fig:2dcut_heatmap} shows the correlations for combined knob cuts, where each knob is cut to the depth that individually achieves the performance target of ~50\% baseline, same as in Figure~\ref{fig:halfipc_compactplot_combined}. 
To quantify the interactions between the pairs, we define a Correlation Score (CS):
\begin{equation}
  \text{CS}(a,b) = \log\!\left(\frac{P_{ab} \cdot P_{base}}{P_a \cdot P_b}\right)
  \label{eq:cs}
\end{equation}

where $P_{base}$ is the baseline performance, $P_a$ and $P_b$ are the performance under two knob's cut respectively, and $P_{ab}$ is the performance under the joint cut. If the two knobs are orthogonal, their combined effect will satisfy
\begin{equation}
\frac{P_{ab}}{P_{base} }
= \frac{P_a}{P_{base}} \cdot \frac{P_b}{P_{base}}
\end{equation}

which yields $\text{CS} = 0$. 

Therefore, a negative CS (blue) indicates \textit{overlapping} behavior, where the joint 
slowdown is less than the product of the independent slowdown, meaning that one knob's bottleneck will mask out the other's performance sensitivity. A positive CS (red) indicates \textit{amplifying} behavior, where the joint slowdown exceeds the product of the independent slowdown, and limiting one bottleneck will exacerbate the other bottleneck. 

Most knob pairs have overlapping behavior, because when one knob already creates a dominant bottleneck, cutting a second knob that targets similar bottlenecks yields diminishing impact. However, a few interesting amplifying pairs stand out. 

In \textit{prefill}, l2 associativity cut paired with DRAM (latency and clock frequency) shows strong amplifying behavior. Large-batch GEMM tiles rely on big working sets that can reside in the L2 and are shared across the SMs, whereas reducing L2 associativity (and thus L2 size) reduces the size of working sets and increases DRAM traffic. Any simultaneous DRAM-side throttling then meets this extra pressure which results in superlinear performance degradation. 

In \textit{decode}, cuts to the DRAM clock and buswidth (dually cutting DRAM bandwidth) exhibit amplified performance degradation. This contrasts with DRAM latency increase paired with either DRAM clock frequency or buswidth cut. Analyzing the source of stalls in each case reveals that when DRAM bandwidth is dually cut, requests stall excessively in the DRAM, whereas cuts to DRAM buswidth and latency cause stalls at the L2 level. In particular, this may suggest that bandwidth cuts paired with latency increases have a dampened impact due to the increased latency reducing pressure on peak requests serviced per unit time. 

\section{Collateral System Impact Analysis}


\subsection{Bottleneck Signature with $\mu$Arch Metrics}
\begin{figure*}[t]
  \centering
  \includegraphics[width=0.99\textwidth, trim=0cm 0.3cm 0cm 0.1cm, clip]{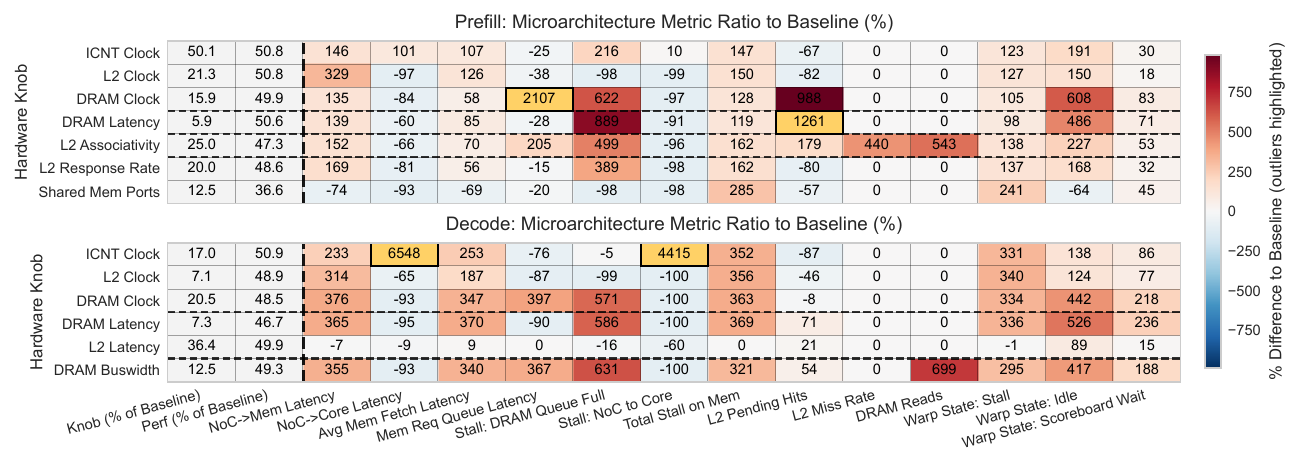}
  \caption{Simulated microarchitecture statistics when cutting each individual knob to achieve closest to 50\% performance target (corresponds to Figure~\ref{fig:halfipc_compactplot_combined}). Different knobs have distinct system bottleneck signatures; a row with with blue and light red cells indicates a ``good'' knob, a shallow cut with the least collateral damage (fewest increases to metric values) to the overall system.}
  \label{fig:uarch_metrics_halfipc}
\end{figure*}

We first dive into microarchitecture metrics for each knob cut. Figure~\ref{fig:uarch_metrics_halfipc} shows the changes of microarchitectural bottleneck metrics collected from AccelSim reports, in \% difference relative to the baseline value when each hardware knob is individually cut to achieve approximately 50\% of original performance (corresponding to Figure~\ref{fig:halfipc_compactplot_combined}). Each row is one knob; the first two columns report the knob's resource cut depth (percentage of original resource availability) and resulting performance. 
Figure~\ref{fig:uarch_metrics_halfipc} shows the following:
\begin{itemize}
    \item A red cell indicates a worse bottleneck (increased pressure) and a blue cell indicates alleviated bottleneck; each knob's row is its unique \emph{bottleneck signature}.
    \item Resource cuts in one place might lead to reduced pressure elsewhere in the system such as less queueing delay with more sparse requests, leading to the blue cells.
    \item Good knobs are those that achieve the performance target with shallower cut and minimal system impact: rows of blue or light red cells, no extreme outliers. They create minimal collateral damage and can enable finer-grained control when targeting intermediate performance levels.
\end{itemize}



\begin{figure*}[t]
  \centering
  \subfloat[Cutting L2 size increases miss rate, which leads to increased DRAM accesses and more core stalls waiting for the memory pipeline.]{\includegraphics[width=0.44\textwidth]{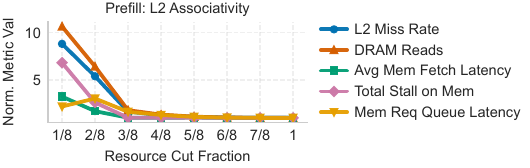}\label{fig:uarch_metrics_Prefill_l2assoc}}
  \hspace{0.03\textwidth}
  \subfloat[Adding L2 latency increases memory-fetch latency and core stalls on loads, while reducing system backpressure in upstream and downstream.]{\includegraphics[width=0.44\textwidth]{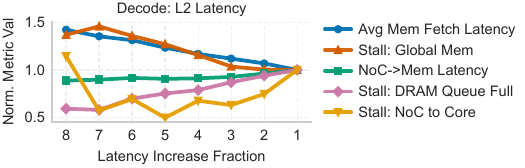}\label{fig:uarch_metrics_Decode_l2_rop_latency}}
  \hfill
  \subfloat[Limiting L2 response rate increases memory roundtrip latency and stalls in memory, and creates high backpressure on DRAM queue.]{\includegraphics[width=0.44\textwidth]{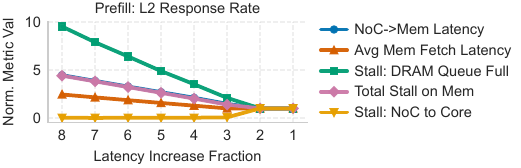}\label{fig:uarch_metrics_Prefill_l2_resp_period}}
  \hspace{0.03\textwidth}
  \subfloat[Reducing accesses to shared memory bank ports greately increases memory stalls and lessens pressure on other parts of the memory system.]{\includegraphics[width=0.44\textwidth]{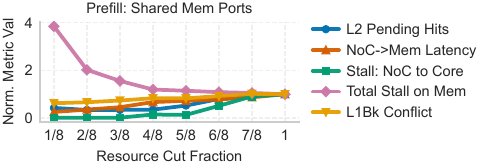}\label{fig:uarch_metrics_Prefill_shmemnumbanks}}
  \caption{Simulated microarchitecture statistics that show each individual knob's collateral impact on the rest of the system.}
  \label{fig:uarch_metrics_halfipc_ind}
\end{figure*}

\textbf{Prefill:}
\emph{Shared Memory Ports} stand out as the cleanest knob for prefill: it creates a purely compute-side bottleneck within the execution core, and has near-zero impact (or even a slightly decrease) on all memory hierarchy bottlenecks, leaving the memory system available for other consumers. It also enables per-core throttling while non-throttled cores can use the memory subsystem.
\emph{L2 Response Rate} and \emph{L2 Associativity} are the next best choices. Reducing L2 response rate raises pressure on DRAM queue occupancy and average fetch latency, but leaves compute-side and NoC metrics largely unaffected. Reducing L2 associativity impacts the cache hierarchy (L2 miss rate and DRAM reads) but not too much of the wider pipeline. Interconnect and L2 clock frequency are also reasonable options, but DRAM-side knobs produce worse system-wide stress and should be avoided when collateral damage is a concern.

\textbf{Decode:}
\emph{L2 Latency} is the most surgical knob for decode: with the shallowest cut ratio, it reaches the performance target while nearly every other metric stays at baseline. Increased L2 latency directly stalls warps in the instruction queue (Warp State: Idle) without impacting DRAM pressure, NoC congestion, or miss rates. This makes L2 Latency the ideal choice with minimal collateral damage. 

\subsection{System Impact of Selected Knobs}

We now examine the microarchitectural metrics that explain why each knob cut degrades performance, and its system-wide effects.

\subsubsection{L2 Size (Prefill)}


We reduce the effective L2 size by reducing associativity (the number of ways) as shown in Figure~\ref{fig:reduce_l2_ways}, which increases the conflicts within each set. For prefill, smaller L2 limits the size of working sets that can fit, and more lines are repeatedly evicted and re-fetched. Figure~\ref{fig:uarch_metrics_Prefill_l2assoc} shows a significant increase in L2 miss rate and DRAM reads at reduced associativity, which increases the total number of core stalls waiting for global memory loads.

As more accesses go through the DRAM, the average memory-fetch latency also increases since it captures the entire round trip of SM~$\rightarrow$~NoC$\rightarrow$~L2/DRAM~$\rightarrow$~NoC$\rightarrow$~SM. The memory-request queue latency rises as well but drops at the most extreme cut, since the system becomes so heavily stalled that requests are temporally spread out with less queue contention.

\subsubsection{L2 Latency (Decode)}


\begin{figure*}[t]
  \centering
  \includegraphics[width=\textwidth]{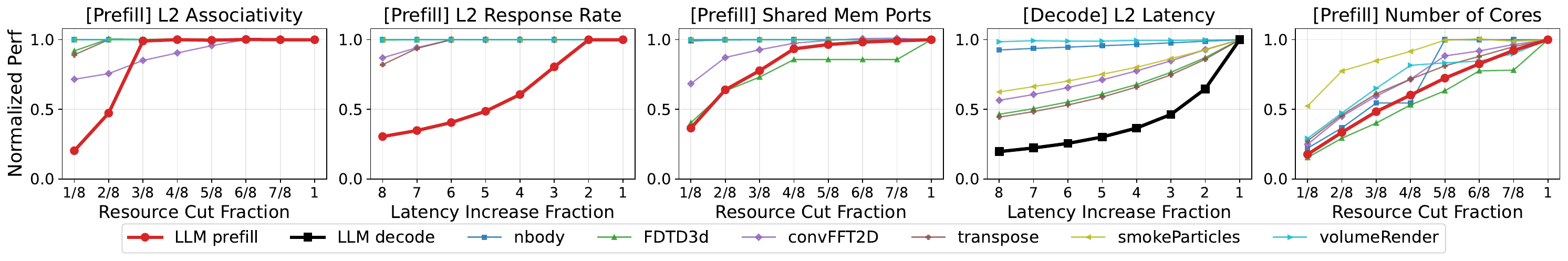}
  \caption{Workload selectivity of the four proposed memory knobs (left panels) versus cutting compute cores (rightmost panel), run on AccelSim with a modeled A100. Bold curves are the targeted LLM kernels; thin curves are six non-LLM workloads. Perf is normalized to each workload's own baseline. The memory knobs throttle the targeted LLM kernel without heavily affecting the others, whereas cutting cores degrades every workload together and shows less selectivity.}
  \label{fig:mwsweep}
\end{figure*}

We increase L2 latency by inserting a fixed delay from the NoC to the L2, as shown in Figure~\ref{fig:latency_throttler}.
The L2 latency is exposed in decode as it is memory-heavy, which also spreads traffic out and reduces system congestion. 

Figure~\ref{fig:uarch_metrics_Decode_l2_rop_latency} shows: (1)~the average memory-fetch roundtrip latency rises, and the number of core stalls on memory loads increases. (2)~Frontend and backend pressure reduces: the NoC$\rightarrow$memory latency (measured from NoC injection to request reaching memory sub-partition) and stalls on full DRAM queue both decrease, as does the stall time attributed to NoC$\rightarrow$SM.



Increasing L2 latency is a low side-effect throttling knob for decode since it throttles decode's critical memory bottleneck without occupying the pipeline and the upstream/downstream components, leaving headroom in system resources for other concurrent workloads. In contrast, for prefill workloads, increased L2 latency also increases the average memory-fetch latency and global-memory stalls, but performance is largely unaffected because the scheduler can draw from a larger pool of ready warps to keep the issue pipelines full and effectively hide the longer L2 latency.

\subsubsection{L2 Bandwidth (Prefill)}


Limiting the response rate of L2 throttles available L2 bandwidth as shown in Figure~\ref{fig:rate_limiter}. L2 bandwidth throttling degrades key metrics relatively linearly, as shown in Figure~\ref{fig:uarch_metrics_Prefill_l2_resp_period}. Reduction in response rate proportionally increases the average memory-fetch latency and the total number of core stalls on global memory loads. Backpressure builds up on the DRAM queue as L2 can no longer process responses fast enough, which in turn increases NoC$\rightarrow$memory latency. In contrast, stalls on the NoC$\rightarrow$core path drop as the pressure is relieved.

\subsubsection{Shared memory BW (Prefill)}

Cutting the shared memory bandwidth via the number of banks available (Figure~\ref{fig:uarch_shmem}) has a large effect on total memory stalls. For prefill, Figure~\ref{fig:uarch_metrics_Prefill_shmemnumbanks} shows a non-linear increase in the total stalls on memory when banks are cut by 4-8x up to peak of 4x the original stall count. Increased pressure at shared memory relieves pressure elsewhere in the memory system. Specifically, the L2 cache and L1D cache see modest drops in bank conflicts and pending hits, respectively, while parts of the system farther out, such as the the interconnect moving data between DRAM and the cores, see a more extreme and rapid decrease.


\subsection{Sensitivity on non-LLM Workloads}
\label{sec:multiwkld}

A throttling knob that caps a targeted AI workload should not render the chip useless for everything else. We show that the proposed knobs are \emph{selective}: a deep cut on targeted LLM kernel has less impact on other workloads. 
We apply each knob on six common non-LLM GPU workloads from CUDA Samples~\cite{cuda_samples}, spanning workload domains of scientific computing, format conversion, and graphics. All workloads use provided default configurations. 

The first four memory-knob panels of Figure~\ref{fig:mwsweep} show that the non-LLM workloads are throttled far less than the targeted LLM kernel. Three L2 knobs have high selectivity, and the shared memory port knob is mostly selective, with one exception, FDTD3d, a stencil that also depends on shared memory.

Compute-side throttling lacks this selectivity. The simplest compute cut, disabling cores, degrades nearly every workload together besides the targeted prefill (rightmost panel of Figure~\ref{fig:mwsweep}).
Throttling compute more surgically, such as cutting issue width or stalling individual function-unit pipelines, requires redesigning the internal core microarchitecture, including issue, scheduling, and fetch logic, which is far more invasive and costly than the proposed memory knobs that switch in a single cycle and reuse existing mechanisms.

\section{End-to-End Performance and Generalization}
\label{sec:e2e_perf}

We now study hot our single-kernel findings generalize to end-to-end inference. Inference runtime is dominated by GEMMs (the $QKV$, output, and FFN projections), with attention secondary. These kernels are \emph{tiled} and the hardware's unit of work has fixed shape. Scheduling optimizations such as dynamic batching change how many tiles run and how full each is packed, not the per-tile shape the hardware sees, thus the throttling sensitivity remains unaffected. Two factors do change the hardware-visible shape, and we analyze each: (i)~in decode attention, continuous batching~\cite{yu2022orca} introduces variable KV cache lengths (Section~\ref{sec:e2e_subsection}), and (ii)~in GEMMs, the $M$ dimension changes with sequence length or batch size (Section~\ref{sec:batch_sweep}). Both have minimal impact on overall trends.

\subsection{Generalization Across Kernel Types}
\label{sec:e2e_subsection}
\subsubsection{Setup}

To show impact of cuts on different kernel types and operator dimensions, we analyze dominant operators in Transformer blocks: $QKV$ projection, self-attention ($QK^V$ and $AV$), output projection, and FFN up and down layers. We construct an estimation by aggregating the runtimes of the five kernels set to the shapes and sizes of three models: DeepSeek-V3~\cite{liu2024deepseek}, Llama-3-70B~\cite{grattafiori2024llama3herdmodels}, and Mixtral-8x7B~\cite{jiang2024mixtral}. We chose not to include element-wise kernels and inter-kernel effects since they typically have minimal runtime as measured in production serving systems~\cite{kwon2023efficient}. 
Runtimes are simulated with AccelSim~\cite{khairy2020accel} on an NVIDIA A100 like configuration~\cite{nvidia_a100_2020}\footnote{For extremely long-running prefill GEMM kernels, we simulated the first $2\times10^9$ instructions and scale cycles to full kernel length by instruction count. 
This matches full-kernel IPC within $<5\%$.}.
Kernels are from the CUTLASS library's high-performance implementations~\cite{nvidia2017cutlass} with parameters from Table~\ref{tab:e2e_config}.

GEMMs use the Stream-K kernel and attention uses the fused FlashAttention kernel~\cite{nvidia2017cutlass, dao2022flashattention, dao2023flashattention, shah2024flashattention}.
For prefill attention, we model the dense multi-head, compute-bound attention where all tokens share the full sequence length (fixed KV). For decode attention, we model the grouped-query attention (GQA)~\cite{ainslie2023gqa} with 8 KV heads for Llama-3 and Mixtral (matching production), where KV-cache traffic dominates the memory-bound kernel. Decode attention uses randomized, various per-request KV lengths (max 4096, mean 2048) to model continuous batching. Without implementation of DeepSeek-V3's multi-head latent attention (MLA)~\cite{liu2024deepseek} in CUTLASS, we approximate it by roofline scaling: compressed latent KV reads ${\sim}3.6\times$ fewer bytes per token than the 8-head GQA kernel which still stays in the same memory-bound regime.

\begin{table}[t]
  \centering
  \caption{Workload configurations for end-to-end aggregation. }
  \label{tab:e2e_config}
  \small
  \begin{tabular}{@{}l l rrr rrr@{}}
    \toprule
    \textbf{Model} & \textbf{Phase} &
    $d$ & $d_{\mathrm{ff}}$ & $h$ &
    $\mathrm{b}$ & $\mathrm{seq}_q$ & $\mathrm{seq}_{kv}$ \\
    \midrule
    \multirow{2}{*}{DeepSeek-V3~\cite{liu2024deepseek}}
      & prefill & \multirow{2}{*}{7168} & \multirow{2}{*}{18432} & \multirow{2}{*}{128}
      & 1   & 4096 & 4096 \\
      & decode  & & & & 128 & 1 & 4096 \\
    \midrule
    \multirow{2}{*}{Llama-3-70B~\cite{grattafiori2024llama3herdmodels}}
      & prefill & \multirow{2}{*}{8192} & \multirow{2}{*}{28672} & \multirow{2}{*}{64}
      & 1   & 4096 & 4096 \\
      & decode  & & & & 128 & 1 & 4096 \\
    \midrule
    \multirow{2}{*}{Mixtral-8x7B~\cite{jiang2024mixtral}}
      & prefill & \multirow{2}{*}{4096} & \multirow{2}{*}{14336} & \multirow{2}{*}{32}
      & 1   & 4096 & 4096 \\
      & decode  & & & & 128 & 1 & 4096 \\
    \bottomrule
  \end{tabular}
    \\[0.6ex]
  {\footnotesize
    $d$: hidden dim; $d_{\mathrm{ff}}$: FFN width (assume SwiGLU-style);
    $h$: number of heads;
    $\mathrm{b}$: batch size;
    $\mathrm{seq}_q$, $\mathrm{seq}_{kv}$: query and KV lengths.
    Llama-3 and Mixtral use GQA~\cite{ainslie2023gqa} and DeepSeek-V3 uses MLA~\cite{liu2024deepseek} for attention in production.}
\end{table}

\begin{figure}[t]
  \centering
  \includegraphics[width=0.49\textwidth, trim=0.1cm 0cm 0cm 0.1cm, clip]{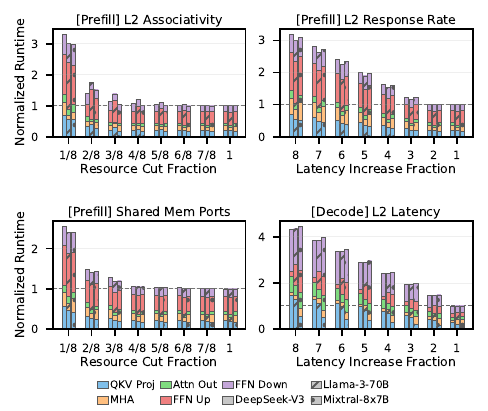}
  {\footnotesize
    In each bar group: left bar--DeepSeek-V3, middle--Llama-3-70B, right--Mixtral-8x7B.}
  \caption{Aggregated end-to-end runtime for LLM models at different resource cut ratios, simulated with AccelSim on a simulated A100 configuration. Per-kernel and overall performance sensitivity aligns with the two representative kernels (Figure~\ref{fig:ipc_vs_resource_both}).}
  \label{fig:e2e_bar}
  
\end{figure}


\subsubsection{Knob effects across different GEMM and attention kernels.}

Figure~\ref{fig:e2e_bar} shows the aggregated runtime for selected knobs and models, where overall performance trends match those in Figure~\ref{fig:ipc_vs_resource_both}. 
All GEMM kernels have nearly identical sensitivity to the resource cuts for prefill and decode respectively, across all three models. The prefill attention kernels follow a similar trend as the GEMMs.
While decode attention is far less sensitive to L2 latency than decode GEMMs, attention is only a small fraction of aggregate runtime, and the GEMMs carry the performance drop. Further analysis on microarchitectural metrics shows that decode attention is deeply DRAM-bandwidth-bound with high data movement along the batch$\times$head dimension with low cache reuse. Across max KV lengths of 1024--4096 (mean 512--2048), its cycle count scales linearly as the KV cache grows across decode steps.

\begin{figure}[t]
  \centering
  \includegraphics[width=0.4\textwidth, trim=0cm 0.4cm 0cm 0.5cm, clip]{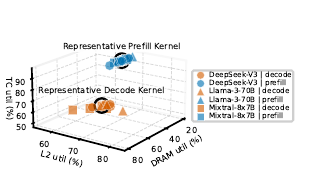}
  \caption{Prefill vs.\ decode GEMM resource utilization on Tensor Core, L2 and DRAM (Nsight Compute, A100). Selected representative kernels for analyzing are circled in black.}
  \label{fig:prefill_decode_gemm_clusters}
\end{figure}

\subsubsection{Representative kernel selection for analysis in depth.}
Among all GEMM kernels, we selected one each for prefill and decode phases for detailed analysis. Figure~\ref{fig:prefill_decode_gemm_clusters} shows the utilization metrics collected on all GEMM kernels across models on a real NVIDIA A100 with Nsight Compute~\cite{nsight-compute}. Prefill and decode GEMMs form two stable clusters;
we select the two kernels circled in Figure~\ref{fig:prefill_decode_gemm_clusters}.


\subsection{Sensitivity Across Batch Sizes}
\label{sec:batch_sweep}
\begin{figure}[t]
  \centering
  \includegraphics[width=0.46\textwidth]{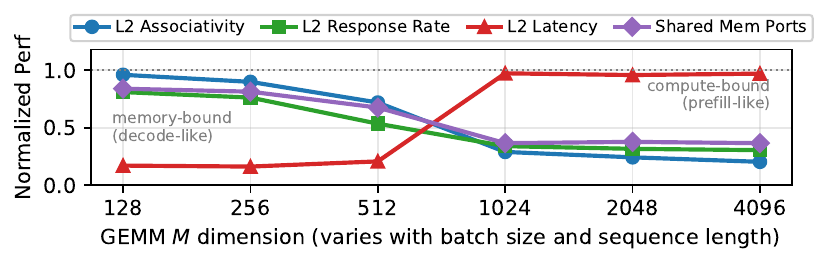}
  \caption{Performance at the deepest ($1/8$) cut of each knob for Mixtral $QKV$ shape with varied GEMM $M$ dimension, representing mixed requests of prefill and decode with varied batch size or sequence length. A lower value means the knob is more effective. The two kernels analyzed in the paper (decode at $M{=}128$, prefill at $M{=}4096$) are the endpoints.}
  \label{fig:per_batch_sweep}
\end{figure}
The GEMM $M$ dimension depends on sequence length (prefill) or batch size (decode), which can change with continuous batching, mixing prefill and decode, or varying request lengths at runtime. To cover these regimes, we sweep $M$ from 128 to 4096 at the fixed $QKV$ shape ($N{=}12288$, $K{=}4096$, mimicking Mixtral) and measure the performance at each knob's 1/8 cut (Figure~\ref{fig:per_batch_sweep}).

As $M$ grows, the kernel transitions from memory-bound (decode-like) to compute-bound (prefill-like): the decode knob (L2 latency) is effective at small $M$, while the prefill knobs (L2 capacity, bandwidth, and shared-memory ports) are effective at large $M$, crossing around $M \in [512,1024]$. Mixed batching requests with both prefill and decode requests fall on this curve at different positions and the throttle effects stay predictable, where multi-knob coupling~\ref{sec:multi_knob_coupling} can be used to find the optimal throttling strategy.

\subsection{Generalization Across Architectures}
\label{sec:new_gens_of_arch}

While our quantitative experiments are on A100-like configurations, the proposed knobs throttle fundamental memory subsystem resources that are common to nearly all GPU architectures regardless of vendor or generation, and we expect the trends to hold. Within NVIDIA's roadmap specifically, the Tensor Memory Accelerator (TMA) introduced since Hopper~\cite{h100_whitepaper} optimizes data movement to Tensor Cores but still utilizes the memory hierarchy, so the L2 throttling mechanisms remain effective. Shared memory size increases across generations, but the bank structure and access patterns are similar, so bank arbitration throttling still applies. The exact sensitivity curves and relative knob rankings may shift across configurations, but as both AI models and newer GPUs rely increasingly on the memory subsystem, memory-side throttling should remain effective and potentially more impactful.

\subsection{Circumvention Resistance}
\label{sec:circumvention_resistance}
Our throttling imposes physical resource caps that software optimization cannot circumvent. Per our threat model (Section~\ref{sec:threat_model}), the constraints are not exposed to software and cannot be bypassed even if an AI observes the performance drop. A capable model could try to rewrite its kernels for the reduced resources, but the CUTLASS Stream-K and FlashAttention kernels we use already self-adapt through tiling and load balancing; after a cut, the throttled resource becomes the dominant bottleneck and the remaining optimization headroom is marginal. In the extreme, the AI could switch to a simpler model, which is precisely the intended effect: a forced reduction in capability.
\section{Related Work}
\label{sec:related_work}

Existing work on resource partitioning/provisioning, traffic shaping, and clock frequency scaling presents similar architectural mechanisms as we propose for bandwidth, size, and latency throttling, but with different purposes. While they target performance optimization, power management or security guarantees, we target AI capability enforcement, but intentionally choose to build on existing mechanisms to reduce design cost. 

Regarding resource provisioning, NVIDIA's Multi-Instance GPU (MIG)~\cite{nvidia_mig_2023} enables hardware-level partitioning of SM cores, cache and memory partitions for workload isolation in multi-tenant environments. ~\cite{li2022miso} studies optimal MIG partition selections for co-located workloads, showing that distinct resource combinations disproportionately affect different workloads. However, MIG operates at a coarse, static level, whereas our mechanisms enable finer-grained resource reshaping during kernel execution. 
The Intel Cache Allocation Technology (CAT)~\cite{intel_cat_2015} allocates LLC ways to different cores to reduce cache contention and prioritize latency-sensitive workloads in a multi-threaded environment. To mitigate poorly distributed cache requests, Sullivan et al.~\cite{per_bank_bandwidth_regulation} regulates bandwidth at a per-bank level by inserting a regulator between cores and a shared cache, which disallows further access to the bank until a set refresh period expires. MemGuard~\cite{memguard} similarly manages memory bandwidth by allocating a per-core bandwidth budget which may be redistributed to other cores based on per-core predictors of expected bandwidth usage.

Traffic shaping mechanisms include Camouflage~\cite{camouflage} which uses a credit-based scheme plus fake requests/responses to shape the inter-arrival time distribution of memory requests and responses, which hides the true distribution from memory side-channel attacks. Traffic shaping more broadly addresses flow control on a network; to handle bursty traffic, Wu et al.~\cite{leaky_bucket} proposes a token bucket scheme to complement the leaky bucket scheme. Both require tokens before sending data, but leaky buckets discard requests when tokens are depleted to enforce a strictly constant rate, while token buckets maintains a pool of tokens to service bursts.

Inserting deliberate delays into the circuit can also be done as a control mechanism.
For example, Sweeney et al.~\cite{logic_locking} adds a variable number of latches to combinational circuits as part of a logic locking scheme to introduce corruption into timing logic.
Dynamic Voltage and Frequency Scaling (DVFS)~\cite{le2010dynamic} is a power saving technique that adjusts the chip's voltage and domain clock frequency based on workload demands. GPU DVFS has been widely studied for power-performance tradeoffs ~\cite{mei2013measurement, mei2017survey} as GPU's power cost is significant.

We make a deliberate design choice to build microarchitectural throttlers from well-established architectural mechanisms to minimize design complexity. We leverage a credit-based rate limiter for L2 bandwidth as in Camouflage~\cite{camouflage}, and add a layer of regulation between shared memory banks and main cores to manage bandwidth similar to ~\cite{per_bank_bandwidth_regulation}. Inserting L2 latency resembles logic locking, and we adapt insights from Intel CAT~\cite{intel_cat_2015} for L2 size throttling. While not explicitly discussed in the paper, potential clock frequency throttling can be implemented in a way similar to DVFS. 

\section{Conclusion}
In this paper, we present the first concrete microarchitectural implementations for dynamic AI performance throttling to defend against malicious AI threats. Through evaluation of ten candidate knobs across the GPU memory hierarchy, we identify four highly effective ones: L2 size (associativity), latency, bandwidth, and shared memory port access rate, that achieve large performance degradation. We propose lightweight hardware implementations similar to existing architectural primitives to minimize design, verification and deployment costs. We show that these knobs achieve the performance targets quickly and reliably at runtime with low collateral impact on the rest of the system. This work fills the gap in hardware-level runtime AI control that has been recognized but unimplemented, with practical mechanisms for GPU vendors to adopt.

\begin{acks}

The authors gratefully acknowledge financial support from the Princeton Laboratory for Artificial Intelligence’s AI$^2$ initiative.   This research was supported in part by a Princeton SEAS Innovation Award.

We thank the members of the Princeton Parallel Group for their insightful discussions. 

We used generative AI tools (Claude Code, Cursor) to assist in batched job scripting and plot visualization, and used Claude and Gemini for editing and language refinement.
\end{acks}


\bibliographystyle{ACM-Reference-Format}
\bibliography{sample-base}

\end{document}